%% file: PrecodedBeam.tex
\documentclass[journal,12pt,onecolumn]{IEEEtran}

\usepackage{setspace}
\ifCLASSOPTIONonecolumn \doublespace \fi
\usepackage{graphics}
\usepackage{rotating}
\usepackage{amsfonts}
\usepackage{amsmath}
\usepackage{multirow}
\usepackage[tight,footnotesize]{subfigure}
\allowdisplaybreaks[1]
\newtheorem{theorem}{Theorem}
\renewcommand{\IEEEQED}{\IEEEQEDopen}

\graphicspath{{Figures/}} \DeclareGraphicsExtensions{.eps,.ps}

\begin{document}

\title{Constellation Precoded Beamforming}

\author{\IEEEauthorblockN{Hong Ju Park and Ender Ayanoglu}\\
\IEEEauthorblockA{Center for Pervasive Communications and Computing\\
Department of Electrical Engineering and Computer Science\\
University of California, Irvine\\
Email: hjpark@uci.edu, ayanoglu@uci.edu}}

\maketitle

\ifCLASSOPTIONonecolumn
 \setlength\arraycolsep{4pt}
\else
 \setlength\arraycolsep{2pt}
\fi

\input{abstract}
\input{introduction}
\input{system_model}
\input{analysis}
\input{design}
\input{simulation}
\input{conclusion}

\bibliographystyle{IEEEtran}
\bibliography{PrecodedBeam.bbl}

\end{document}

%% file: abstract.tex
\begin{abstract}
We present and analyze the performance of constellation precoded
beamforming. This multi-input multi-output transmission technique is
based on the singular value decomposition of a channel matrix. In
this work, the beamformer is precoded to improve its diversity
performance. It was shown previously that while single beamforming
achieves full diversity without channel coding, multiple beamforming
results in diversity loss. In this paper, we show that a properly
designed constellation precoder makes uncoded multiple beamforming
achieve full diversity order. We also show that partially precoded
multiple beamforming gets better diversity order than multiple
beamforming without constellation precoder if the subchannels to be
precoded are properly chosen. We propose several criteria to design
the constellation precoder. Simulation results match the analysis,
and show that precoded multiple beamforming actually outperforms
single beamforming without precoding at the same system data rate
while achieving full diversity order.
\end{abstract}

%% file: introduction.tex
\section{Introduction} \label{sec:introduction}

When the perfect channel state information is available at the
transmitter, beamforming is employed to achieve spatial multiplexing
and thereby increase the data rate, or to enhance the performance of
a multi-input multi-output (MIMO) system \cite{jafarkhaniBook}. The
beamforming vectors are designed in \cite{SampathJCOM01},
\cite{palomarTSP03} for various design criteria, and can be obtained
by singular value decomposition (SVD), leading to a
channel-diagonalizing structure optimum in minimizing the average
bit error rate (BER) \cite{palomarTSP03}. Uncoded single
beamforming, which carries only one symbol at a time, was shown to
achieve full diversity order of $MN$ where $M$ is the number of
receive antennas and $N$ is the number of transmit antennas
\cite{sengulTC06AnalSingleMultpleBeam}, \cite{OrdonezTSP07}.
However, uncoded multiple beamforming, which increases the
throughput by sending multiple symbols at a time, loses full
diversity order over flat fading channels
\cite{sengulTC06AnalSingleMultpleBeam}, \cite{OrdonezTSP07}.

It is known that an SVD subchannel with larger singular value
provides larger diversity gain. During the simultaneous parallel
transmission of the symbols on the diagonalized subchannels, the
performance is dominated by the subchannel with the smallest
singular value. To overcome the degradation of the diversity order
of multiple beamforming, bit-interleaved coded multiple beamforming
(BICMB) was proposed \cite{akayTC06BICMB},
\cite{akayTC06BICMB_arxiv}. This scheme interleaves the codewords
through the multiple subchannels with different diversity order,
resulting in better diversity order. BICMB can achieve the full
diversity order offered by the channel as long as the code rate
$R_c$ and the number of subchannels used $S$ satisfy the condition
$R_c S \leq 1$ \cite{Park_arxiv_0809_5096}, \cite{ParkITAW09}. In
this paper, we present an uncoded single and multiple beamforming
technique that achieves full diversity order. This technique employs
the constellation precoding scheme \cite{GamalJIT03},
\cite{XinJWCOM03}, \cite{LiuJCOM03}, \cite{ZhangJCOM07},
\cite{GressetGlobecom09}, which is used for space-time or
space-frequency block codes to increase the system data rate without
losing full diversity order. We show via analysis and simulations
that fully precoded multiple beamforming achieves full diversity
order even in the absence of any channel coding. For this purpose,
we derive an upper bound for the pairwise error probability of the
precoded beamforming system. Several criteria to design the
precoding matrix are proposed. Simulation results show that fully
precoded multiple beamforming with a properly designed precoding
matrix outperforms single beamforming at the same system data rate.

The rest of this paper is organized as follows. The description of
precoded beamforming is given in Section \ref{sec:system_model}.
Section \ref{sec:analysis} calculates the upper bound of pairwise
error probability for different schemes of precoded beamforming. In
Section \ref{sec:precoder_design}, several criteria to design the
precoding matrix are proposed. Simulation results supporting the
analysis are shown in Section \ref{sec:simulation}. Finally, we end
the paper with a conclusion in Section \ref{sec:conclusion}.

%% file: system_model.tex
\section{System Model} \label{sec:system_model}

The MIMO channel $\mathbf{H} \in \mathbb{C}^{M \times N}$ is assumed
to be quasi-static, Rayleigh, and flat fading, and perfectly known
to both the transmitter and the receiver. The beamforming vectors
are determined by the SVD of the MIMO channel, i.e., $\mathbf{H} =
\mathbf{U \Lambda V^H}$ where $\mathbf{U}$ and $\mathbf{V}$ are
unitary matrices, and $\mathbf{\Lambda}$ is a diagonal matrix whose
$s^{th}$ diagonal element, $\lambda_s \in \mathbb{R}$, is a singular
value of $\mathbf{H}$ in decreasing order. When $S$ symbols are
transmitted at the same time, then the first $S$ vectors of
$\mathbf{U}$ and $\mathbf{V}$ are chosen to be used as beamforming
matrices at the receiver and the transmitter, respectively. In Fig.
\ref{fig:structure_CPM} which displays the structure of
constellation precoded beamforming, $\mathbf{\tilde{U}}$ and
$\mathbf{\tilde{V}}$ denote the beamforming matrices picked from
$\mathbf{U}$ and $\mathbf{V}$. Depending on $S$ and the number of
symbols precoded $R$, constellation precoded beamforming can be
classified into three types as are described below.
\ifCLASSOPTIONonecolumn
\begin{figure}[!m]
\centering{ \subfigure[Precoded single
beamforming]{\includegraphics[width=0.75\linewidth]{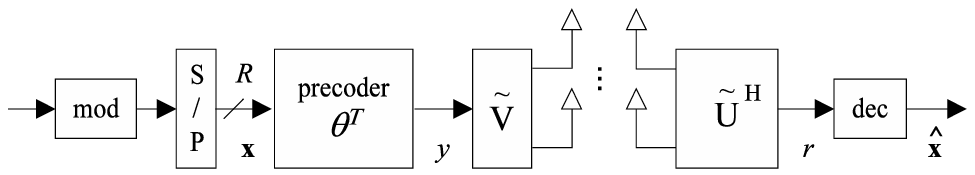}
\label{fig:structure_PSB}} \hfil \subfigure[Precoded multiple
beamforming]{\includegraphics[width=0.75\linewidth]{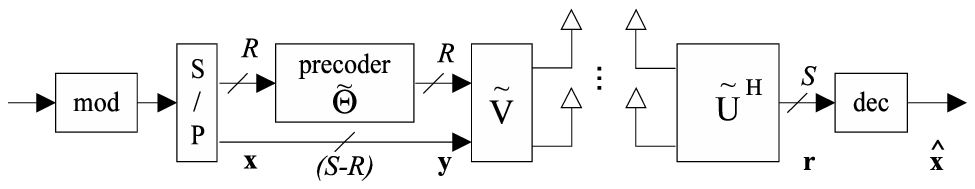}
\label{fig:structure_PMB}}} \caption{Structure of Constellation
Precoded Beamforming.} \label{fig:structure_CPM}
\end{figure}
\else
\begin{figure}[!t]
\centering{ \subfigure[Precoded single
beamforming]{\includegraphics[width=1\linewidth]{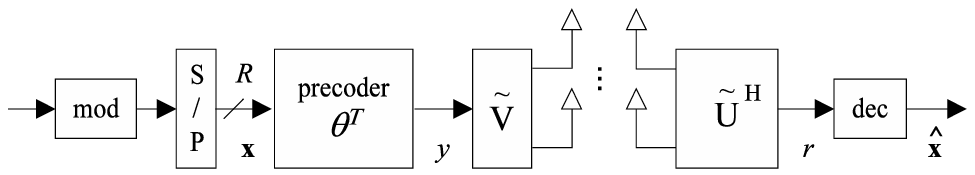}
\label{fig:structure_PSB}} \hfil \subfigure[Precoded multiple
beamforming]{\includegraphics[width=1\linewidth]{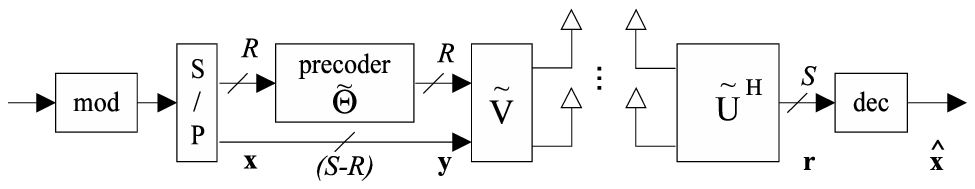}
\label{fig:structure_PMB}}} \caption{Structure of Constellation
Precoded Beamforming.} \label{fig:structure_CPM}
\end{figure}
\fi

\subsection{Precoded Single Beamforming} \label{sec:Model_PSB}

In the Precoded Single Beamforming (PSB) scheme, a precoder
$\boldsymbol{\theta}^T$ generates a symbol $y$ from an $R \times 1$
modulated symbol vector $\mathbf{x}$, $y = \boldsymbol{\theta}^T
\mathbf{x}$. We assume that each of the $R$ members of $\mathbf{x}$
belongs to a normalized signal set $\chi \subset \mathbb{C}$ of size
$|\chi| = 2^m$, such as $2^m$-QAM, where $m$ is the number of input
bits to the Gray encoder. Due to the beamforming vectors employed at
both of the ends, the precoded symbol is transmitted on the
subchannel with the largest singular value. Hence, the detected
symbol $r$ is written as
\begin{align}
r = \lambda_1 \boldsymbol{\theta}^T \mathbf{x} + n
\label{eq:deteced_PSB}
\end{align}
where $\lambda_1$ is the largest singular value of the channel
matrix $\mathbf{H}$, and $n$ is an additive white Gaussian noise
with zero mean and variance $N_0 = N / SNR$. $\mathbf{H}$ is complex
Gaussian with zero mean and unit variance, and to make the received
signal-to-noise ratio $SNR$, the total transmitted power is scaled
as $N$. The maximum likelihood (ML) decoding of the detected symbol
is given by
\begin{align}
\mathbf{\hat{x}} = \min_{\mathbf{x} \in \chi^R} | r - \lambda_1
\boldsymbol{\theta}^T \mathbf{x} |^2. \label{eq:MLdecoding_PSB}
\end{align}
The system data rate for PSB is $\eta = m \cdot R$ bits/channel use.

\subsection{Precoded Multiple Beamforming} \label{sec:Model_PMB}

In this scheme, $S>1$ modulated symbols are simultaneously
transmitted on the subchannels with the largest $S$ singular values.
The $S \times 1$ symbol vector $\mathbf{x}$ whose elements belong to
$\chi$ are precoded by a square precoding matrix $\mathbf{\Theta}$.
The constellation precoder can be expressed as
\begin{align}
\mathbf{\Theta} = \mathbf{P} \times \left[ \begin{array}{cc}
\mathbf{\tilde{\Theta}} & \mathbf{0} \\
\mathbf{0} & \mathbf{I}_{S-R}
\end{array} \right]
\label{eq:precoder_def}
\end{align}
where $\mathbf{\tilde{\Theta}}$ is $R \times R$ constellation
precoding matrix that precodes the first $R$ modulated symbols of
the vector $\mathbf{x}$, and $\mathbf{P}$ is an \mbox{$S \times S$}
permutation matrix to define the mapping of the precoded and
non-precoded symbols onto the predefined subchannels. When all of
the $S$ modulated symbols are precoded ($R = S$), we call the
resulting system Fully Precoded Multiple Beamforming (FPMB),
otherwise, we call it Partially Precoded Multiple Beamforming
(PPMB). For FPMB, $\mathbf{P}$ can be replaced by the identity
matrix, and $\mathbf{\Theta} = \mathbf{\tilde{\Theta}}$. The $S
\times 1$ detected symbol vector $\mathbf{r}$ at the receiver is
written as
\begin{align}
\mathbf{r} = \mathbf{\Lambda}_S \mathbf{\Theta x} + \mathbf{n}
\label{eq:deteced_PMB}
\end{align}
where $\mathbf{\Lambda}_S$ is a diagonal matrix whose elements are
the first $S$ singular values of $\mathbf{\Lambda}$, and
$\mathbf{n}$ is an additive white Gaussian noise vector. The ML
decoding of the detected symbol is given by
\begin{align}
\mathbf{\hat{x}} = \min_{\mathbf{x} \in \chi^S} \left\| \mathbf{r} -
\mathbf{\Lambda}_S \mathbf{\Theta x} \right\|^2.
\label{eq:MLdecoding_PMB}
\end{align}
The system data rate for precoded multiple beamforming is $\eta = m
\cdot S$ bits/channel use.

%% file: analysis.tex
\section{Diversity Analysis} \label{sec:analysis}

In this section, we will calculate the diversity order by analyzing
the pairwise error probability (PEP) between the transmitted symbol
$\mathbf{x} = \left[ x_1 \, \dots \, x_S \right]^T$ and the detected
symbol $\mathbf{\hat{x}} = \left[ \hat{x}_1 \, \dots \, \hat{x}_S
\right]^T$.

\subsection{Precoded Single Beamforming} \label{sec:PSB}

For the ML decoding criterion of (\ref{eq:MLdecoding_PSB}), the
instantaneous PEP can be expressed as
\begin{align}
\mathrm{Pr} \left( \mathbf{x} \rightarrow \mathbf{\hat{x}} \mid
\mathbf{H} \right) &= \mathrm{Pr} \left( | r - \lambda_1
\boldsymbol{\theta}^T \mathbf{x} |^2 \geq | r - \lambda_1
\boldsymbol{\theta}^T \mathbf{\hat{x}} |^2 \mid
\mathbf{H} \right) \nonumber \\
&= \mathrm{Pr} \left( \beta \geq | \lambda_1 \boldsymbol{\theta}^T
(\mathbf{x} - \mathbf{\hat{x}}) | ^2 \mid \mathbf{H} \right)
\label{eq:instantaneous_PEP_PSB}
\end{align}
where $\beta = -\lambda_1 [ \boldsymbol{\theta}^T (\mathbf{x} -
\mathbf{\hat{x}}) ]^* n - \lambda_1 n^*  \boldsymbol{\theta}^T
(\mathbf{x} - \mathbf{\hat{x}})$. Since $\beta$ is a zero mean
Gaussian random variable with variance \mbox{$2 N_0 \lambda_1^2 |
\boldsymbol{\theta}^T (\mathbf{x} - \mathbf{\hat{x}}) | ^2$},
(\ref{eq:instantaneous_PEP_PSB}) is rewritten as
\begin{align}
\mathrm{Pr} \left( \mathbf{x} \rightarrow \mathbf{\hat{x}} \mid
\mathbf{H} \right) = Q \left( \sqrt{\frac{\lambda_1^2 |
\boldsymbol{\theta}^T (\mathbf{x} - \mathbf{\hat{x}}) |^2}{2 N_0}}
\right) \label{eq:instantaneous_PEP2_PSB}
\end{align}
where $Q(\cdot)$ is the well-known $Q$ function. By using the upper
bound on the $Q$ function $Q(x) \leq \frac{1}{2} e^{-x^2/2}$, the
average PEP can be expressed as
\begin{align}
\mathrm{Pr} \left( \mathbf{x} \rightarrow \mathbf{\hat{x}} \right)
&= E \left[ \mathrm{Pr} \left( \mathbf{x} \rightarrow
\mathbf{\hat{x}} \mid \mathbf{H}
\right) \right] \nonumber \\
&\leq E \left[ \frac{1}{2} \exp \left(- \frac{\lambda_1^2 |
\boldsymbol{\theta}^T (\mathbf{x} - \mathbf{\hat{x}}) |^2}{4 N_0}
\right) \right]. \label{eq:PEP_PSB}
\end{align}
Assume that $| \boldsymbol{\theta}^T ( \mathbf{x} - \mathbf{\hat{x}}
) |^2 > 0$ for a distinct pair of $\mathbf{x}$ and
$\mathbf{\hat{x}}$. Previously, in
\cite{sengulTC06AnalSingleMultpleBeam}, \cite{Park_arxiv_0809_5096},
and \cite{ParkITAW09}, we showed the closed form expression of
(\ref{eq:PEP_PSB}). We provide a formal description of the result
from \cite{Park_arxiv_0809_5096}, \cite{ParkITAW09} below.
\begin{theorem}
Consider the largest \mbox{$S \leq \min(M, N)$} eigenvalues $\mu_s$
of the uncorrelated central \mbox{$M \times N$} Wishart matrix that
are sorted in decreasing order, and a weight vector
\mbox{$\boldsymbol{\alpha} = [\alpha_1 \, \cdots \, \alpha_S]^T$}
with non-negative real elements. In the high signal-to-noise ratio
regime, an upper bound for the expression $E [ \exp (-\gamma
\sum_{s=1}^S \alpha_s \mu_s ) ]$ which is used in the diversity
analysis of a number of MIMO systems is
\begin{align}
E\left[ \exp \left( - \gamma \sum\limits_{s=1}^S \alpha_s \mu_s
\right) \right] \leq \zeta \left( \alpha_{min} \gamma
\right)^{-(M-\delta+1)(N-\delta+1)}
\end{align}
where $\gamma$ is signal-to-noise ratio, $\zeta$ is a constant,
$\alpha_{min} = \min \{ \alpha_1, \, \cdots, \, \alpha_S \}$, and
$\delta$ is the index to the first non-zero element in the weight
vector. \label{theorem:E_PEP}
\end{theorem}
\begin{IEEEproof}
See \cite{Park_arxiv_0809_5096}, \cite{ParkITAW09}.
\end{IEEEproof}

Applying Theorem \ref{theorem:E_PEP} to (\ref{eq:PEP_PSB}) where
$S=1$, $\delta = 1$, and  $\alpha_{min} = |\boldsymbol{\theta}^T (
\mathbf{x} - \mathbf{\hat{x}} ) | ^2$, we get the upper bound for
the PEP as
\begin{align}
\mathrm{Pr} \left( \mathbf{x} \rightarrow \mathbf{\hat{x}} \right)
&\leq \frac{1}{2} \left( \frac{| \boldsymbol{\theta}^T ( \mathbf{x}
- \mathbf{\hat{x}} ) | ^2}{4 N} SNR \right)^{-MN}.
\label{eq:PEP_PSB_final}
\end{align}
Therefore, it is easily found that PSB achieves full diversity order
of $MN$ once it satisfies the condition
\begin{align}
| \boldsymbol{\theta}^T ( \mathbf{x} - \mathbf{\hat{x}} ) |^2 > 0
\label{eq:condition_full_diversity}
\end{align}
for any distinct pair of $\mathbf{x}$ and $\mathbf{\hat{x}}$. The
method to design the precoding vector will be described in Section
\ref{sec:precoder_design}.

\subsection{Fully Precoded Multiple Beamforming} \label{sec:FPMB}

By using the same approach in PSB, we get the upper bound to the
instantaneous PEP for precoded multiple beamforming as
\begin{align}
\mathrm{Pr} \left( \mathbf{x} \rightarrow \mathbf{\hat{x}} \mid
\mathbf{H} \right) &= \mathrm{Pr} \left( \left\| \mathbf{r} -
\mathbf{\Lambda}_S \mathbf{\Theta x} \right\|^2 \geq
\left\|\mathbf{r} - \mathbf{\Lambda}_S \mathbf{\Theta \hat{x}}
\right\|^2 \mid
\mathbf{H} \right) \nonumber \\
& \leq \frac{1}{2} \exp \left(- \frac{\left\| \mathbf{\Lambda}_S
\mathbf{\Theta} (\mathbf{x} - \mathbf{\hat{x}})\right\|^2}{4 N_0}
\right).
\label{eq:instantaneous_PEP_PMB}
\end{align}
Let's define $\mathbf{d} = \left[d_1 \, \cdots \, d_S \right] ^T$ as
a Euclidean vector that results from the precoded symbols of a
distinct pair $\mathbf{x}$ and $\mathbf{\hat{x}}$. Then, $\mathbf{d}
= \mathbf{\Theta} ( \mathbf{x} - \mathbf{\hat{x}} )$, and the
absolute value $|d_i|$ can be interpreted as a Euclidean distance
between the symbols belonging to a new constellation transformed by
the $i^{th}$ row vector of $\mathbf{\Theta}$ from the original
constellation. For FPMB, the average PEP is expressed as
\begin{align}
\mathrm{Pr} \left( \mathbf{x} \rightarrow \mathbf{\hat{x}} \right)
&\leq E \left[ \frac{1}{2} \exp \left(- \frac{\sum\limits_{s=1}^S
\lambda_s^2 | d_s | ^2}{4 N_0} \right) \right]. \label{eq:PEP_FPMB}
\end{align}
Applying Theorem \ref{theorem:E_PEP} to (\ref{eq:PEP_FPMB}), we get
the upper bound to PEP as
\begin{align}
\mathrm{Pr} \left( \mathbf{x} \rightarrow \mathbf{\hat{x}} \right)
&\leq \zeta \left( \frac{\hat{d}_{min}}{4N} SNR
\right)^{-(M-\delta+1)(N-\delta+1)} \label{eq:PEP_FPMB_final}
\end{align}
where $\zeta$ is a constant, $\hat{d}_{min} = \min \{|d_1|^2, \,
\cdots, \, |d_S|^2\}$, and $\delta$ is an index to the first
non-zero element of the (squared) Euclidean distance vector $\left[
|d_1|^2 \, \cdots \, |d_S|^2 \right]$. Therefore, FPMB also achieves
full diversity order if $\delta$ from any distinct pair is equal to
$1$, which implies that $|d_1|^2 = |\boldsymbol{\theta}_1^T
(\mathbf{x} - \mathbf{\hat{x}})|^2
> 0$ for any distinct pair, where $\boldsymbol{\theta}_1^T$ is the first row vector of $\mathbf{\Theta}$. The way to build the precoding matrix
will be described in Section \ref{sec:precoder_design}.

\subsection{Partially Precoded Multiple Beamforming} \label{sec:PPMB}

The partial precoding scheme divides the modulated symbols into two
groups of symbols, i.e., precoded and non-precoded symbols. Through
the permutation and the grouping, the numerator of the exponent term
in (\ref{eq:instantaneous_PEP_PMB}) can be represented as described
below. For this purpose, let's define $\mathbf{b}_p = \left[ b_p(1)
\, \cdots \, b_p(R) \right]$ as a vector whose element $b_p(k)$ is
the subchannel on which the precoded symbols are transmitted, and
$b_p(k) < b_p(l)$ for $k < l$. In the same way, $\mathbf{b}_n =
\left[ b_n(1) \, \cdots \, b_n(S-R) \right]$ is defined as an
increasingly ordered vector whose element $b_n(k)$ is the subchannel
which carries the non-precoded symbols. By reordering the resulting
vector $\mathbf{\Lambda}_S \mathbf{\Theta} (\mathbf{x} -
\mathbf{\hat{x}})$ for a simpler representation of
$\mathbf{\Lambda}_S \mathbf{P}$, the numerator of the exponent term
in (\ref{eq:instantaneous_PEP_PMB}) can be expressed as
\begin{align}
\left\| \mathbf{\Lambda}_S \mathbf{\Theta} (\mathbf{x} -
\mathbf{\hat{x}}) \right\| &= \left\| \mathbf{\Lambda}_S \mathbf{P}
\left[
\begin{array}{cc}
\mathbf{\tilde{\Theta}} & \mathbf{0} \\
\mathbf{0} & \mathbf{I}_{S-R}
\end{array} \right] (\mathbf{x} - \mathbf{\hat{x}}) \right\| \nonumber \\
&= \left\| \left[ \begin{array}{cc}
\mathbf{\Lambda}_p & \mathbf{0} \\
\mathbf{0} & \mathbf{\Lambda}_n \end{array} \right] \left[
\begin{array}{cc}
\mathbf{\tilde{\Theta}} & \mathbf{0} \\
\mathbf{0} & \mathbf{I}_{S-R}
\end{array} \right] \left[ \begin{array}{c}
\mathbf{x}_p - \mathbf{\hat{x}}_p \\
\mathbf{x}_n - \mathbf{\hat{x}}_n
\end{array} \right] \right\| \nonumber \\
&= \left\| \left[ \begin{array}{c}
\mathbf{\Lambda}_p \mathbf{\tilde{\Theta}} (\mathbf{x}_p - \mathbf{\hat{x}}_p) \\
\mathbf{\Lambda}_n (\mathbf{x}_n - \mathbf{\hat{x}}_n)
\end{array} \right] \right\| \label{eq:numerator}
\end{align}
where $\mathbf{\Lambda}_p$ and $\mathbf{\Lambda}_n$ are the $R
\times R$ and $(S-R) \times (S-R)$ diagonal matrices whose elements
are the ordered singular values corresponding to the subchannels of
the vectors $\mathbf{b}_p$ and $\mathbf{b}_n$, respectively, and
similarly $\mathbf{x}_p = \left[ x_1 \, \cdots \, x_R \right]$,
$\mathbf{x}_n = \left[ x_{R+1} \, \cdots \, x_S \right]$,
$\mathbf{\hat{x}}_p = \left[ \hat{x}_1 \, \cdots \, \hat{x}_R
\right]$, $\mathbf{\hat{x}}_n = \left[ \hat{x}_{R+1} \, \cdots \,
\hat{x}_S \right]$. By plugging (\ref{eq:numerator}) in
(\ref{eq:instantaneous_PEP_PMB}), we get an upper bound to PEP for
PPMB as
\begin{align}
\mathrm{Pr} \left( \mathbf{x} \rightarrow \mathbf{\hat{x}} \right)
&\leq E \left[ \frac{1}{2} \exp \left(- \frac{\kappa}{4 N_0} \right)
\right] \label{eq:PEP_PPMB_equation}
\end{align}
where
\begin{align}
\kappa = \sum\limits_{i = 1}^R \lambda_{b_p(i)}^2 | \tilde{d}_i | ^2
&+ \sum\limits_{i = 1}^{S-R} \lambda_{b_n(i)}^2 | x_{R+i} -
\hat{x}_{R+i} | ^2 \label{eq:PEP_PPMB}
\end{align}
and $\tilde{d}_i$ is the $i^{th}$ element of a Euclidean vector
$\mathbf{\tilde{d}} =$ \mbox{$\mathbf{\tilde{\Theta}} (\mathbf{x}_p
- \mathbf{\hat{x}}_p)$}. Let's assume that the constellation
precoding matrix $\mathbf{\tilde{\Theta}}$ meets the condition of
FPMB to achieve full diversity order. Since (\ref{eq:PEP_PPMB}) has
the closed form expression similar to (\ref{eq:PEP_FPMB_final}) as
described in FPMB, $\delta$ value needs to be obtained from a
composite vector with such kind of elements as $|\tilde{d}_i|^2$ and
$|x_{R+i} - \hat{x}_{R+i}|^2$, to observe the diversity behavior of
a given pairwise error. In addition, a different pair can lead to
different diversity behavior. Therefore, we need to get the maximum
$\delta$ out of all the possible pairwise errors to decide the
diversity order of a given PPMB system.

All of the distinct pairs of $\mathbf{x}$ and $\mathbf{\hat{x}}$ can
be divided into three groups in terms of $\mathbf{x}_p$,
$\mathbf{\hat{x}}_p$, $\mathbf{x}_n$, and $\mathbf{\hat{x}}_n$. The
first group includes the pairs that have \mbox{$\mathbf{x}_p =
\mathbf{\hat{x}}_p$}, and the second group comprises the pairs
satisfying \mbox{$\mathbf{x}_p \neq \mathbf{\hat{x}}_p$}, but
\mbox{$\mathbf{x}_n = \mathbf{\hat{x}}_n$}. Finally, the last group
is consisted of the pairs that \mbox{$\mathbf{x}_p \neq
\mathbf{\hat{x}}_p$}, and \mbox{$\mathbf{x}_n \neq
\mathbf{\hat{x}}_n$}. We will present the method to calculate
$\delta$ for a pair of each group, and to find $\delta_{max}$ for
each group.

Since the vector $\mathbf{\tilde{d}}$ is a zero vector for the first
group, the first summation of $\kappa$ in (\ref{eq:PEP_PPMB}) is
zero, resulting in $\delta$ being equal to the minimum of
$\mathbf{b}_n$. By considering all of the possible pairs, we can
easily see that $b_n(1) \leq \delta \leq b_n(S-R)$. Therefore, the
maximum value is $\delta_1 = b_n(S-R)$ which corresponds to the pair
satisfying $x_i = \hat{x}_i$ for all $i$ except \mbox{$i =
b_n(S-R)$}. For any pair in the second group, the term with the
first singular value survives in $\kappa$, according to the
inherited property of the constellation precoding matrix, i.e.,
$|\tilde{d}_1|^2 > 0$. However, the second summation in $\kappa$
disappears since $\mathbf{x}_n = \mathbf{\hat{x}}_n$. Therefore, the
maximum value of this group $\delta_2 = b_p(1)$. Now, for the third
group, both summations in $\kappa$ exist. Then, $\delta$ is chosen
to be the smaller value between the minimum of $\mathbf{b}_n$ and
$b_p(1)$. In the same manner as was already given in the analysis of
the first group, the maximum of the minimum of $\mathbf{b}_n$ can be
found to be $b_n(S-R)$. Therefore, the maximum $\delta$ for this
group is $\delta_3 = \max \{ b_p(1), \, b_n(S-R) \}$. Finally,
$\delta_{max}$ can be decided as
\begin{align}
\delta_{max} &= \max \{ \delta_1, \, \delta_2, \, \delta_3 \} \nonumber \\
&= \max \{ b_n(S-R), \, b_p(1), \max \{ b_p(1), \, b_n(S-R) \}
\} \nonumber \\
&= \max \{ b_p(1), \, b_n(S-R) \}. \label{eq:Q_max}
\end{align} \\
\textbf{Example:} We provide the diversity analysis of the $4 \times
4$ partially precoded multiple beamforming system with \mbox{$S =
4$} and \mbox{$R = 2$}. In this example, we assume that the precoded
symbols are transmitted on the subchannel $1$ and $3$ while the
non-precoded symbols are transmitted on the subchannel $2$ and $4$.
Then, this configuration gives $\mathbf{b}_p = \left[ 1 \,\, 3
\right]$, and $\mathbf{b}_n = \left[ 2 \,\, 4 \right]$. By following
the result in (\ref{eq:Q_max}), $\delta_{max}$ is equal to $\max
\{1, \, 4 \} = 4$, leading to the diversity order of $1$. The
pairwise errors, satisfying $x_1 = \hat{x}_1, x_2 = \hat{x}_2, x_3 =
\hat{x}_3$, but $x_4 \neq \hat{x}_4$, inflict loss on the diversity
order of this system. Table \ref{tab:partial_order} summarizes the
diversity order analysis for all of the possible combinations of the
$4 \times 4$ partially precoded multiple beamforming system.

\ifCLASSOPTIONonecolumn
\begin{table}[!m]
\else
\begin{table}[!t]
\fi \caption{Diversity order of $4 \times 4$, $S = 4$ partially
precoded multiple beamforming system}
\begin{center}
\begin{tabular}{|c|c|c|c|c|c|c|}
\hline
$R$ & $\mathbf{b}_p$ & $\mathbf{b}_n$ & $b_p(1)$ & $b_n(S-R)$ & $\delta_{max}$ & $O_{div}$\\
\hline \hline
\multirow{6}*{$2$} & $[1 \, 2]$ & $[3 \, 4]$ & $1$ & $4$ & $4$ & $1$ \\
\cline{2-7}
& $[1 \, 3]$ & $[2 \, 4]$ & $1$ & $4$ & $4$ & $1$ \\
\cline{2-7}
& $[1 \, 4]$ & $[2 \, 3]$ & $1$ & $3$ & $3$ & $4$ \\
\cline{2-7}
& $[2 \, 3]$ & $[1 \, 4]$ & $2$ & $4$ & $4$ & $1$ \\
\cline{2-7}
& $[2 \, 4]$ & $[1 \, 3]$ & $2$ & $3$ & $3$ & $4$ \\
\cline{2-7}
& $[3 \, 4]$ & $[1 \, 2]$ & $3$ & $2$ & $3$ & $4$ \\
\hline \hline
\multirow{4}*{$3$} & $[1 \, 2 \, 3]$ & $[4]$ & $1$ & $4$ & $4$ & $1$ \\
\cline{2-7}
& $[1 \, 2 \, 4]$ & $[3]$ & $1$ & $3$ & $3$ & $4$ \\
\cline{2-7}
& $[1 \, 3 \, 4]$ & $[2]$ & $1$ & $2$ & $2$ & $9$ \\
\cline{2-7}
& $[2 \, 3 \, 4]$ & $[1]$ & $2$ & $1$ & $2$ & $9$ \\
\hline
\end{tabular}
\end{center}
\label{tab:partial_order}
\end{table}

%% file: design.tex
\section{Precoder Design} \label{sec:precoder_design}

\subsection{Precoded Single Beamforming} \label{sec:design_Single_beamforming}

The optimum precoding vector should maximize the array gain as well
as meet the condition for achieving full diversity order. Based on
(\ref{eq:PEP_PSB_final}), the optimum precoding vector
$\boldsymbol{\theta}_{opt}$ can be determined by solving the
following equation:
\begin{align}
\boldsymbol{\theta}_{opt} = \arg \max_{\boldsymbol{\theta}}
\min_{\forall \mathbf{x} \neq \mathbf{\hat{x}}} |
\boldsymbol{\theta}^T ( \mathbf{x} - \mathbf{\hat{x}} ) |
\label{eq:optimal_vector}
\end{align}
subject to the power constraint \mbox{$E |
 \boldsymbol{\theta}^T \mathbf{x} | ^2 = E \left\| \mathbf{x} \right\| ^2
= R$}. The problem in (\ref{eq:optimal_vector}) coincides with that
of \cite{DamenThesis} which addressed multi-user coding. In
\cite{DamenThesis}, a number of users are assumed to transmit
simultaneously at the same power, each of whom uses a rotated
version of QAM symbols. Thus, the noiseless received symbol belongs
to a new constellation containing the sum of each rotated QAM
symbols. For a simple construction, the rotation vector is defined
as
\begin{align}
\boldsymbol{\theta} = \left[ 1 \quad e^{j \phi} \quad e^{j2\phi}
\quad e^{j4\phi} \quad \cdots \quad e^{j 2^{R-2}\phi} \right]^T
\label{eq:form_vector}
\end{align}
where $j = \sqrt{-1}$. The optimum rotation angles for $R \leq 7$
are found by searching $\phi$ that maximizes the minimum squared
Euclidean distance of the new constellation. Hence, we can apply the
result of \cite{DamenThesis} to precoded single beamforming.

\subsection{Precoded Multiple Beamforming} \label{sec:design_Multiple_beamforming}

To design the precoding matrix, we establish various design
criteria. Since we focus on even power distribution on each
subchannel, the precoding matrix is restricted to be a unitary
matrix which preserves the power.

\subsubsection{Maximization of the Minimum Euclidean Distance Among the $S$ Coordinates, $\mathbf{\Phi}_1$} \label{sec:criteria_maxarraygain}

This criterion minimizes the upper bound to PEP of
(\ref{eq:PEP_FPMB_final}) by maximizing $\hat{d}_{min}$ as
\begin{align}
\mathbf{\Theta}_{opt} = \arg \max_{\mathbf{\Theta}} \min_{\forall
\mathbf{x} \neq \mathbf{\hat{x}}} \hat{d}_{min}
\label{eq:optimal_matrix_crit_maxcoding}
\end{align}
subject to the power constraint \mbox{$E \|
 \mathbf{\Theta} \mathbf{x} \| ^2 = E \left\| \mathbf{x} \right\| ^2
= R$}. Since the analytical solution of this problem is unavailable,
computer search can be used for small $S$ and small constellation
sizes. For this purpose, we employ the parameterization method of
complex unitary matrices in \cite{XinJWCOM03}. In this method, any
$S \times S$ unitary matrix $\mathbf{\Sigma}$ can be written as
\begin{align}
\mathbf{\Sigma} = \mathbf{D} \prod_{1 \leq k \leq S-1, k+1 \leq l <
S} \mathbf{G}_{kl} \left( \psi_{kl}, \rho_{kl} \right)
\label{eq:parameterization}
\end{align}
where $\mathbf{D}$ is an $S \times S$ diagonal unitary matrix,
$\psi_{kl} \in [ -\pi, \pi]$, $\rho_{kl} \in [ -\pi/2, \pi/2]$, and
$\mathbf{G}_{kl} \left( \psi_{kl}, \rho_{kl} \right)$ is a complex
Givens matrix, which is the $S \times S$ identity matrix with the
$(k,k)^{th}$, $(l,l)^{th}$, $(k,l)^{th}$, $(l,k)^{th}$ elements
substituted by $\cos \psi_{kl}$, $\cos \psi_{kl}$, $e^{-j\rho_{kl}}
\sin \psi_{kl}$, and $-e^{j\rho_{kl}} \sin \psi_{kl}$, respectively.
The optimization of the diagonal unitary matrix $\mathbf{D}$ is not
necessary since $\hat{d}_{min}$ is the squared absolute value of an
element which includes the diagonal entry of $\mathbf{D}$ with the
magnitude equal to one. Therefore, we need to optimize only $S(S-1)$
parameters of $\{\psi_{kl}, \phi_{kl}\}$. The optimum precoding
matrix found by this method automatically satisfies the full
diversity order condition since $|d_1|^2 \geq \hat{d}_{min} > 0$.

\subsubsection{Maximization of the Minimum Euclidean Distance of the First Coordinate, $\mathbf{\Phi}_2$} \label{sec:criteria_maxerror}

For large $S$, the squared Euclidean distance of $i^{th}$ coordinate
$|d_i|^2$ is smaller than $\lambda_1^2$. Thus, the term $\lambda_1^2
|d_1|^2$ takes the biggest portion of the summation in
(\ref{eq:PEP_FPMB}). This fact leads to an idea that maximization of
the minimum Euclidean distance of the first coordinate $|d_1|^2$ can
lower the PEP. In this criterion, we solve the following equation as
\begin{align}
\mathbf{\Theta}_{opt} = \arg \max_{\mathbf{\Theta}} \min_{\forall
\mathbf{x} \neq \mathbf{\hat{x}}} |\boldsymbol{\theta}_1^T (
\mathbf{x} - \mathbf{\hat{x}} )|^2
\label{eq:optimal_matrix_crit_maxdominant}
\end{align}
subject to the power constraint \mbox{$E \|
 \mathbf{\Theta} \mathbf{x} \| ^2 = E \left\| \mathbf{x} \right\| ^2
= R$}. Since it is difficult to solve
(\ref{eq:optimal_matrix_crit_maxdominant}) in a tractable way, and
the optimization equation is the same as (\ref{eq:optimal_vector}),
we propose a method that adopts the result of
(\ref{eq:optimal_vector}). In this method, we use the optimum
precoding vector (\ref{eq:optimal_vector}) in
$\mathbf{\Theta}_{opt}$ as the first row vector. To build a unitary
matrix, we utilize the \mbox{$S$-point} inverse fast Fourier
transform (IFFT) matrix $\mathbf{F}_S$, whose $(l,m)$ element is
given by $\sqrt S \exp(j 2\pi (l-1)(m-1)/S)$. The IFFT matrix
provides two properties we can use; the elements of the first column
vector are all ones, and the IFFT matrix is a unitary matrix. By
constructing a precoding matrix as
\begin{align}
\mathbf{\Theta}_{opt} = \mathbf{F}_S^T \mathrm{diag}(
\boldsymbol{\hat{\theta}}_{opt}  )
\label{eq:optimal_matrix_crit_maxdominant_construct}
\end{align}
where $\boldsymbol{\hat{\theta}}_{opt}$ is the precoding vector
obtained from (\ref{eq:optimal_vector}), we can see that the unitary
matrix $\mathbf{\Theta}_{opt}$ contains
$\boldsymbol{\hat{\theta}}_{opt}$ as the first row vector. This
method guarantees full diversity order since $|d_1|^2 > 0$.

\subsubsection{Maximization of the Geometric Mean, $\mathbf{\Phi}_3$} \label{sec:criteria_maxgeomean}

Since the summation in (\ref{eq:PEP_FPMB}) consists of many terms,
the previous optimizations which optimize only one term may not
necessarily be the best criterion. Maximizing the arithmetic mean or
the geometric mean of $|d_i|^2$ values are attractive candidates
which consider all of $|d_i|^2$ values simultaneously. Between the
two, we choose the geometric mean since the arithmetic mean does not
necessarily guarantee the full diversity order condition. However,
the geometric mean meets the condition on $|d_1|^2 > 0$ since the
geometric mean of the optimum precoding matrix in this sense will be
larger than zero. Therefore, the optimization is based on
\begin{align}
\mathbf{\Theta}_{opt} = \arg \max_{\mathbf{\Theta}} \min_{\forall
\mathbf{x} \neq \mathbf{\hat{x}}} \prod\limits_{i=1}^S
|\boldsymbol{\theta}_i^T ( \mathbf{x} - \mathbf{\hat{x}} )|^{2/S}
\label{eq:optimal_matrix_crit_maxgeomean}
\end{align}
subject to the same power constraint as the previous criteria, and
$\boldsymbol{\theta}_i^T$ is the $i^{th}$ row vector of the
precoding matrix $\mathbf{\Theta}$. Authors in \cite{XinJWCOM03}
introduced an algebraic method to construct the precoding matrix in
the space-time diversity system. According to \cite{XinJWCOM03}, the
unitary precoding matrix can be written as
\begin{align}
\mathbf{\Theta}_{opt} = \mathbf{F}_S^T \mathrm{diag} \left( 1,
\sigma, \cdots, \sigma^{S-1} \right) \label{eq:LCP-B}
\end{align}
where $\sigma = e^{j2\pi/P}$, and the method to determine $P$ is
available in \cite{XinJWCOM03}.

%% file: simulation.tex
\section{Simulation Results} \label{sec:simulation}

To illustrate the analysis of the diversity order in Section
\ref{sec:analysis}, we now present simulation results over various
channel dimensions. Fig. \ref{fig:2x2_sb_psb_fpmb},
\ref{fig:3x3_sb_psb_fpmb}, and \ref{fig:4x4_sb_psb_fpmb} show bit
error rate (BER) performance of SB, PSB, and FPMB. The curves with
legend FPMB $\mathbf{\Phi}_1$, $\mathbf{\Phi}_2$, $\mathbf{\Phi}_3$
are generated by the precoding matrices based on each criterion in
Section \ref{sec:precoder_design}. For a fair comparison between the
different schemes, the system data rate $\eta$ is set to $4$, $6$,
$8$ bits/channel use. Throughout the figures, PSB and FPMB are shown
to achieve full diversity order since the slopes are parallel to
that of SB which is known to achieve full diversity order. A
comparison between SB and PSB reveals that SB outperforms PSB for
any channel dimension. The reason is that the array gain which can
be observed by PEP is related to the minimum squared Euclidean
distance, and the minimum squared Euclidean distance of the new
constellation generated by the precoding vector is smaller than that
of the SB constellation. For example, the minimum squared Euclidean
distance of $\eta = 4$ PSB normalized constellation is $0.27$, while
that of normalized $16$-QAM SB is $0.4$.

Contrary to the case of $2 \times 2$, FPMB outperforms SB for larger
channel dimension. In the case of $3 \times 3$, FPMB
$\mathbf{\Phi}_3$ gives $2$ dB gain over SB. A bigger gain of $6$ dB
is observed by FPMB $\mathbf{\Phi}_3$ for the case of $4 \times 4$.
Bigger gain for larger dimension can be explained by a comparison of
the instantaneous PEPs of SB in
\cite{sengulTC06AnalSingleMultpleBeam} and FPMB in
(\ref{eq:PEP_FPMB}). It can be stated that a larger number of
singular values lead to a bigger array gain. We also find that the
maximization of the geometric mean results in better performance
than the others for larger channel dimension.

\ifCLASSOPTIONonecolumn
\begin{figure}[!m]
\centering \includegraphics[width =
0.75\linewidth]{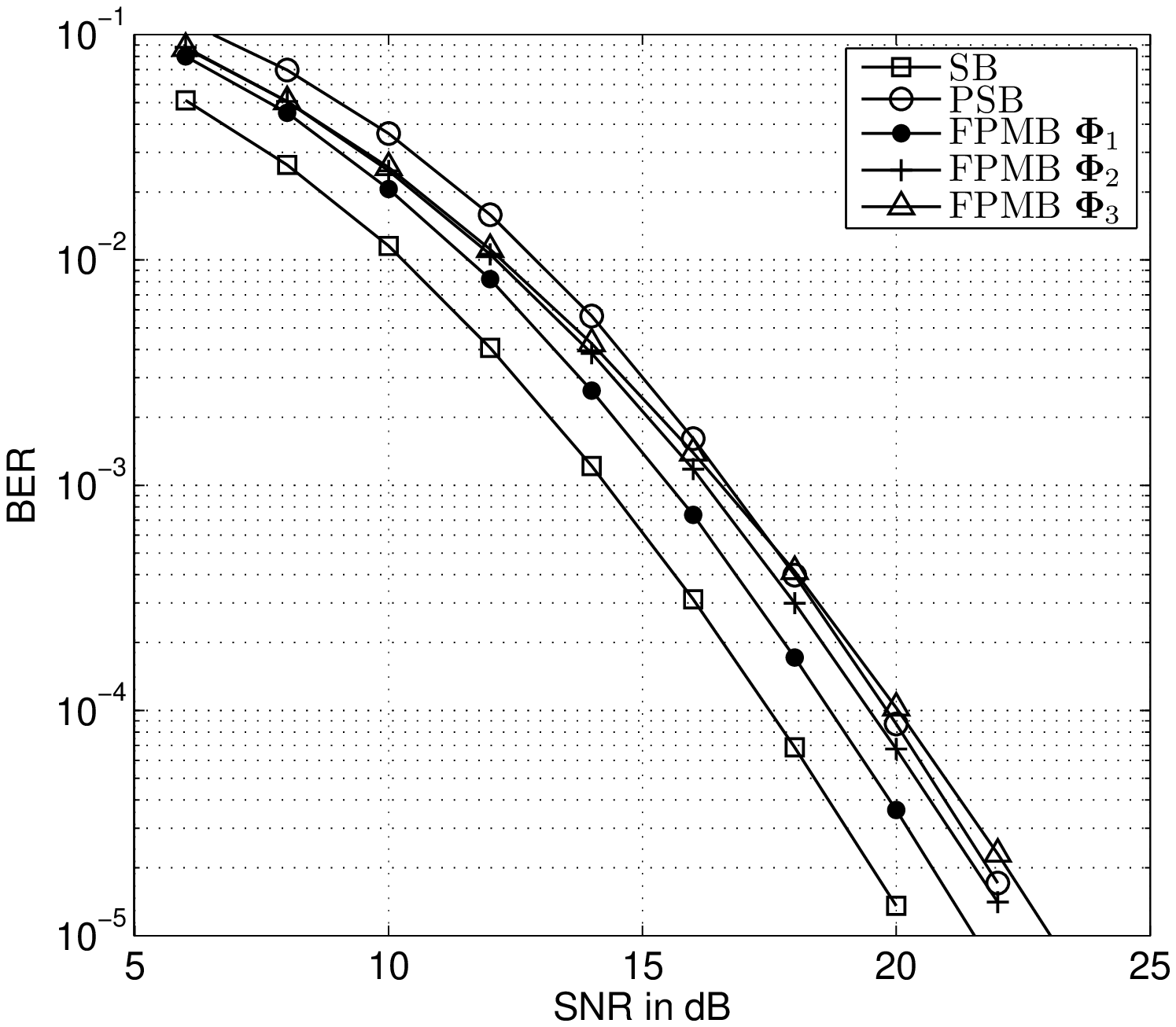} \caption{BER vs. SNR comparison
for $2 \times 2$ $16$-QAM SB, $4$-QAM $R=2$ PSB, and $4$-QAM $S=2$
FPMB.} \label{fig:2x2_sb_psb_fpmb}
\end{figure}
\else
\begin{figure}[!t]
\centering \includegraphics[width =
0.95\linewidth]{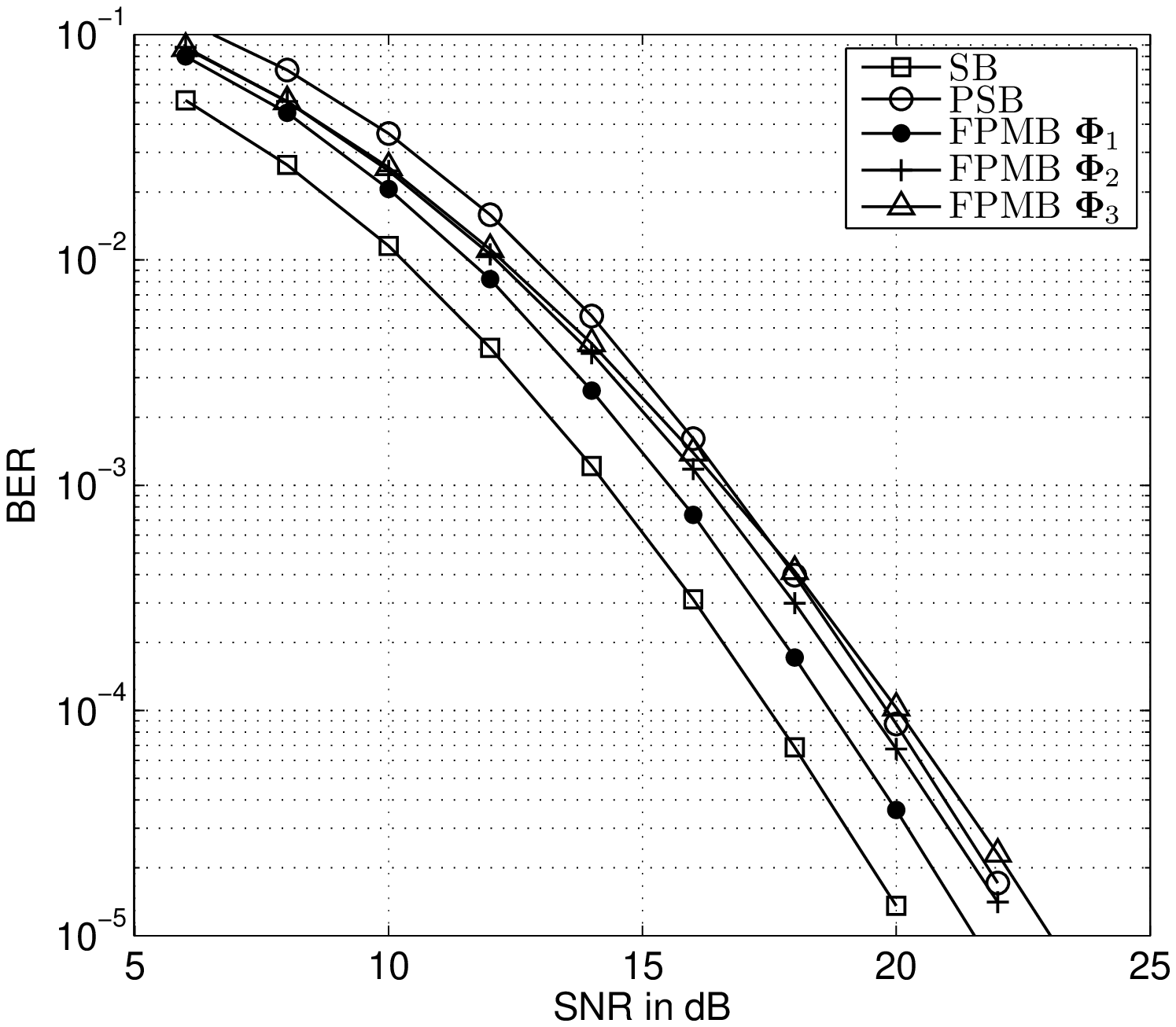} \caption{BER vs. SNR comparison
for $2 \times 2$ $16$-QAM SB, $4$-QAM $R=2$ PSB, and $4$-QAM $S=2$
FPMB.} \label{fig:2x2_sb_psb_fpmb}
\end{figure}
\fi

\ifCLASSOPTIONonecolumn
\begin{figure}[!m]
\centering \includegraphics[width =
0.75\linewidth]{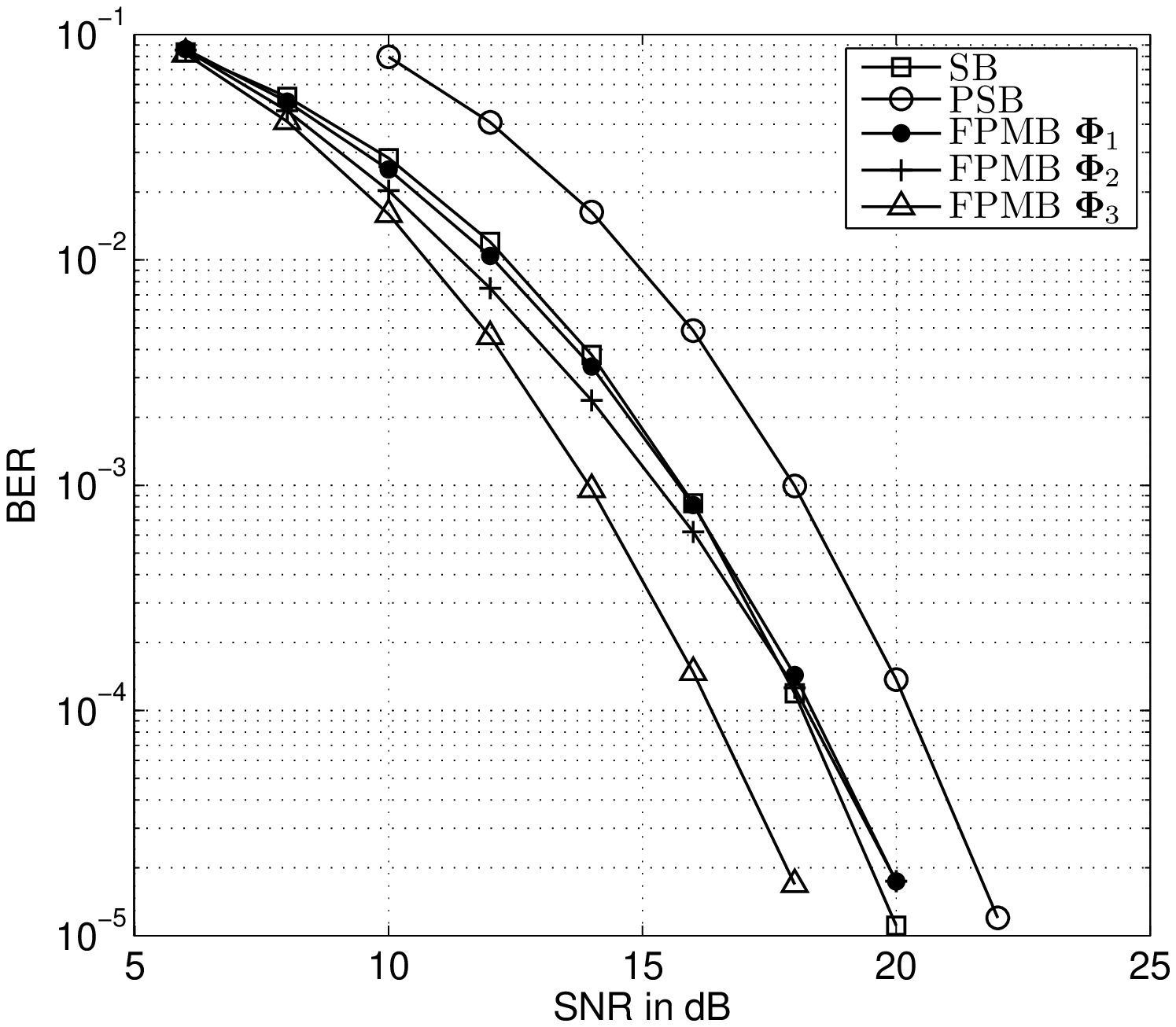} \caption{BER vs. SNR comparison
for $3 \times 3$ $64$-QAM SB, $4$-QAM $R=3$ PSB, and $4$-QAM $S=3$
FPMB.} \label{fig:3x3_sb_psb_fpmb}
\end{figure}
\else
\begin{figure}[!t]
\centering \includegraphics[width =
0.95\linewidth]{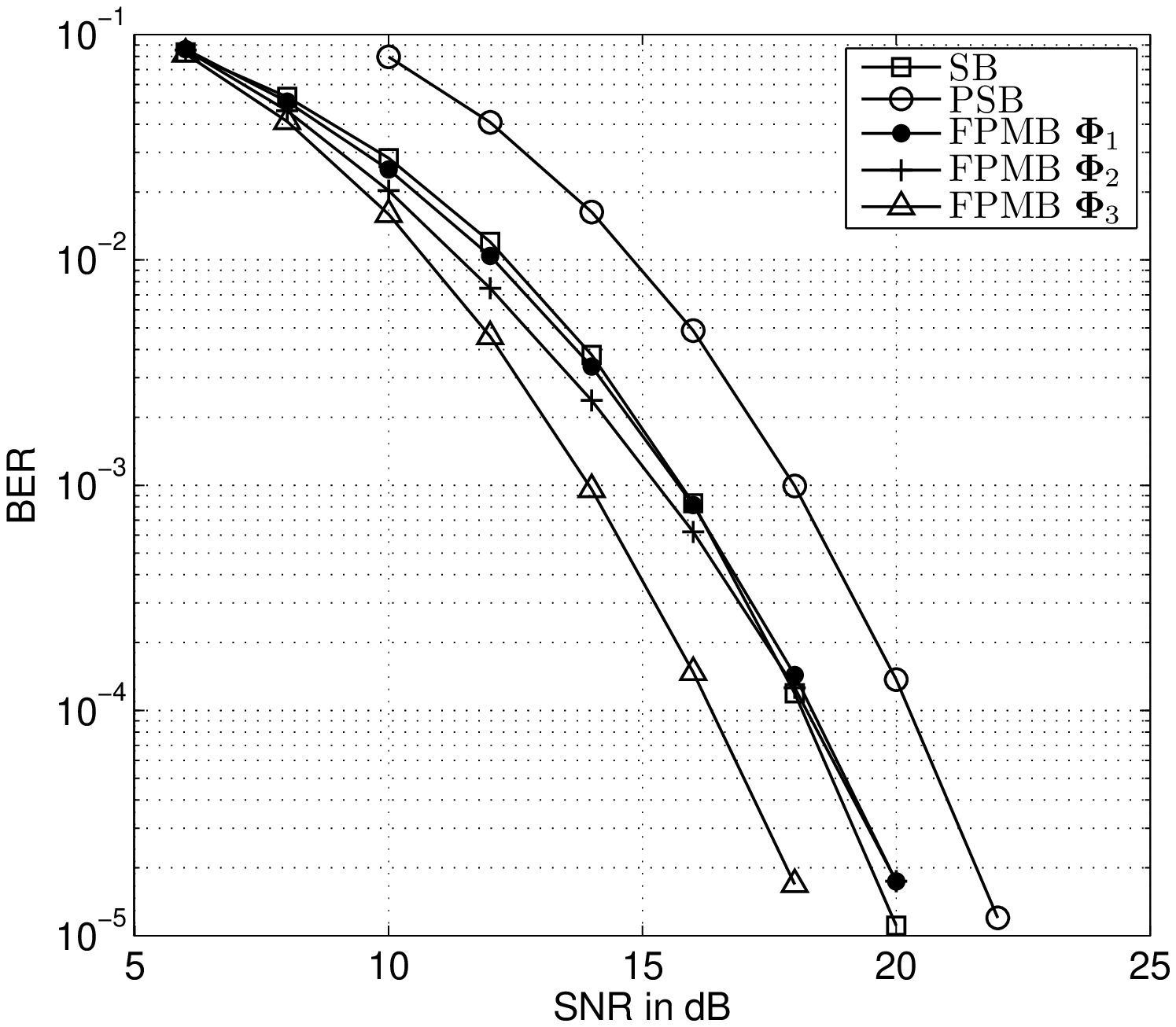} \caption{BER vs. SNR comparison
for $3 \times 3$ $64$-QAM SB, $4$-QAM $R=3$ PSB, and $4$-QAM $S=3$
FPMB.} \label{fig:3x3_sb_psb_fpmb}
\end{figure}
\fi

\ifCLASSOPTIONonecolumn
\begin{figure}[!m]
\centering \includegraphics[width =
0.75\linewidth]{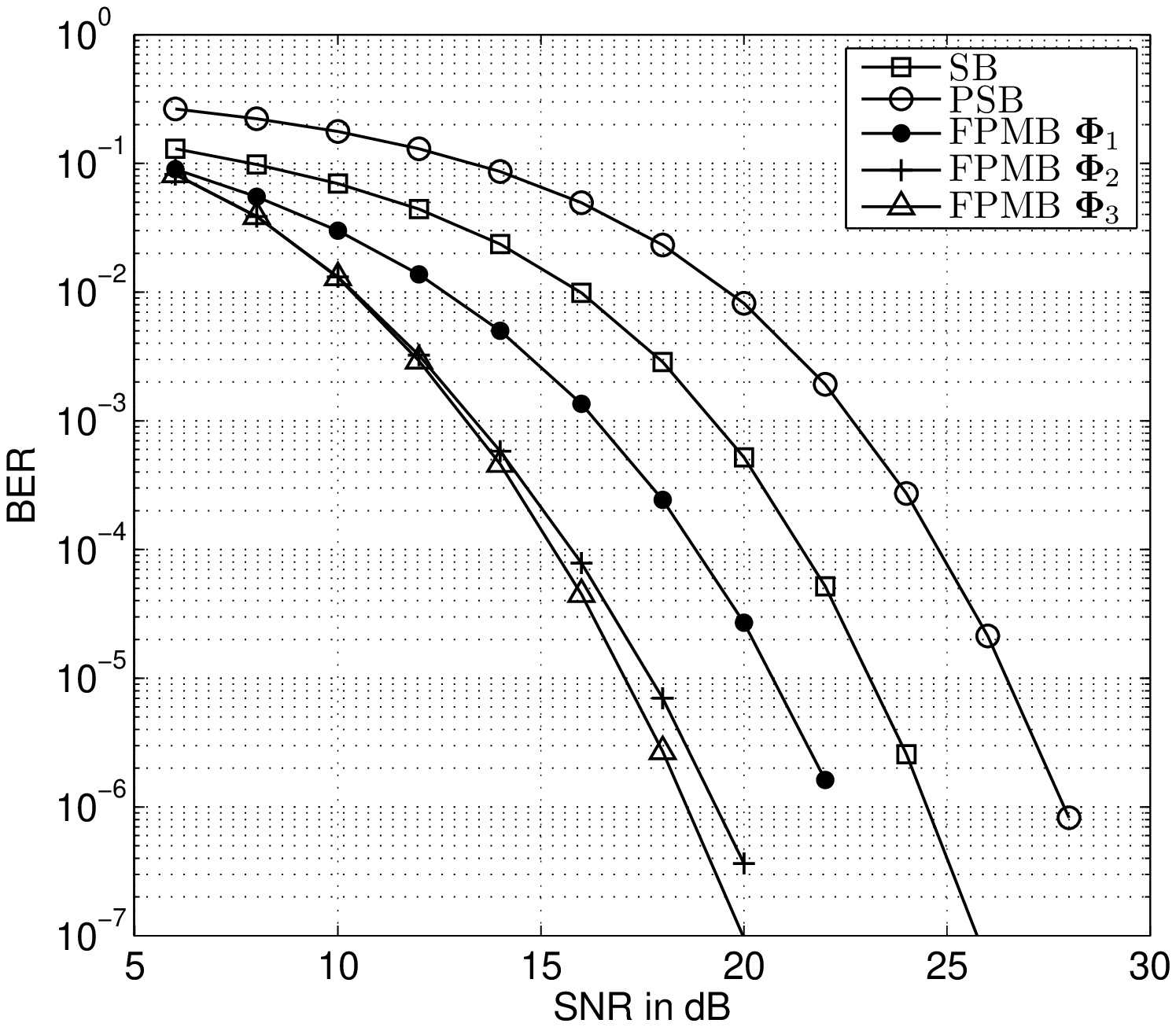} \caption{BER vs. SNR comparison
for $4 \times 4$ $256$-QAM SB, $4$-QAM $R=4$ PSB, and $4$-QAM $S=4$
FPMB.} \label{fig:4x4_sb_psb_fpmb}
\end{figure}
\else
\begin{figure}[!t]
\centering \includegraphics[width =
0.95\linewidth]{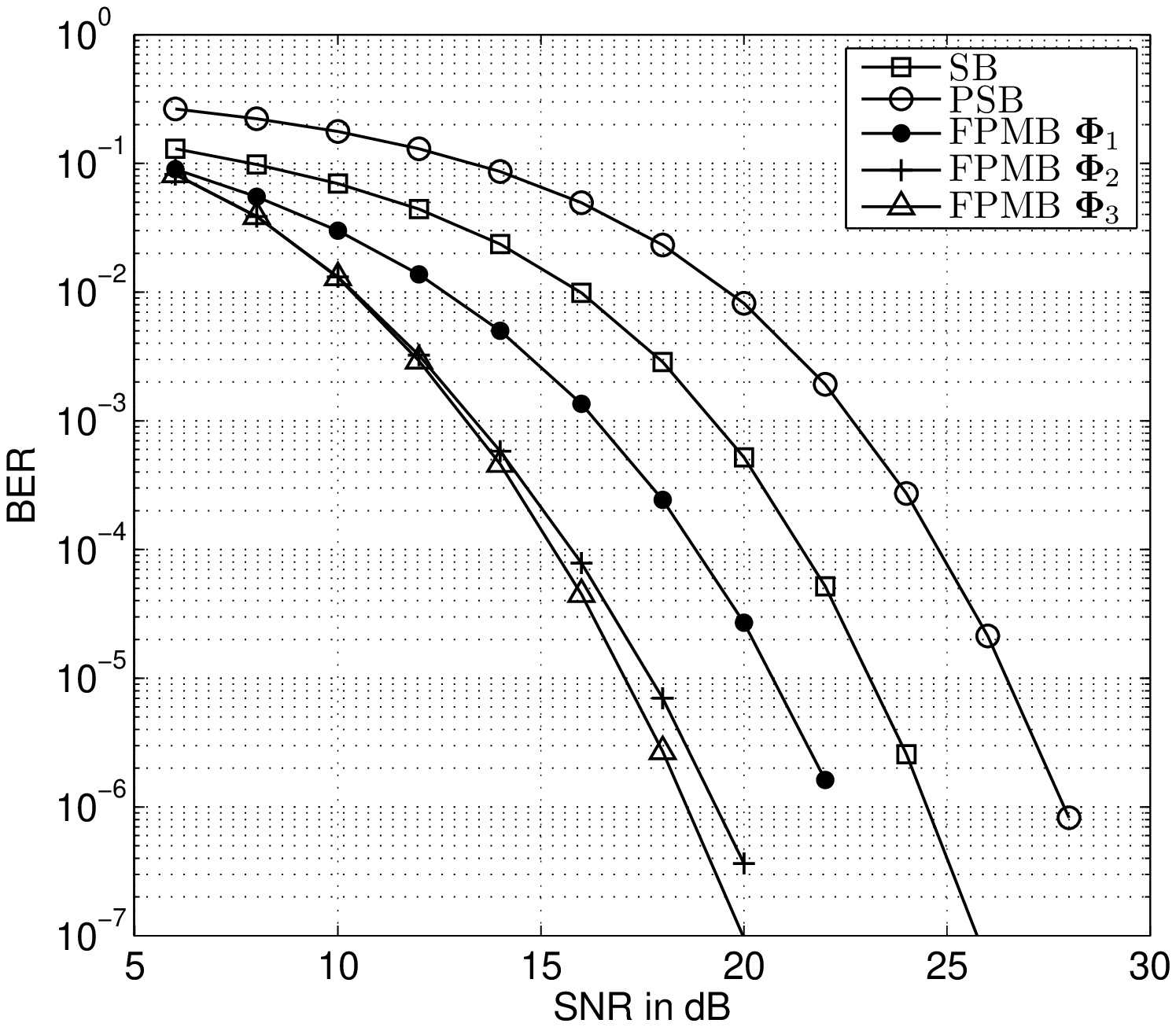} \caption{BER vs. SNR comparison
for $4 \times 4$ $256$-QAM SB, $4$-QAM $R=4$ PSB, and $4$-QAM $S=4$
FPMB.} \label{fig:4x4_sb_psb_fpmb}
\end{figure}
\fi

Simulation results to support the diversity analysis of $4 \times 4$
$S=4$ PPMB in Table \ref{tab:partial_order} are provided in Fig.
\ref{fig:partial_precoding}. We find that the simulation results
follow the diversity orders in Table \ref{tab:partial_order}. The
curves with the same diversity order give different array gain
depending on the subchannel selection of precoded symbol
transmission. BER at high SNR for \mbox{$\mathbf{b}_p = [1\, 4]$},
$[2\, 4]$, $[1\, 2\, 4]$ are the same, leading to around $10$ dB
larger gain than that of \mbox{$\mathbf{b}_p = [3\, 4]$}. This can
be explained by the fact that PPMB with the smaller number of pairs
(causing the worst diversity order) provide larger array gain. Since
the subchannel $3$ transmitting the non-precoded symbol dominates
the performance loss for \mbox{$\mathbf{b}_p = [1\, 4]$}, $[2\, 4]$,
$[1\, 2\, 4]$, only one pair, satisfying \mbox{$x_3 = \hat{x}_3$}
and \mbox{$x_i \neq \hat{x}_i$} for \mbox{$i = 1, 2, 4$}, is related
to the worst diversity order. On the other hand, for
\mbox{$\mathbf{b}_p = [3\, 4]$} where the subchannel $3$ also
dominates the performance loss, half of the total possible pairs,
such that \mbox{$x_1 = \hat{x}_1$}, and \mbox{$x_2 = \hat{x}_2$},
and \mbox{$x_3 \neq \hat{x}_3$}, or \mbox{$x_4 \neq \hat{x}_4$},
make the worst diversity order. This fact also applies to the case
of diversity order of $9$.

\ifCLASSOPTIONonecolumn
\begin{figure}[!m]
\centering \includegraphics[width =
0.75\linewidth]{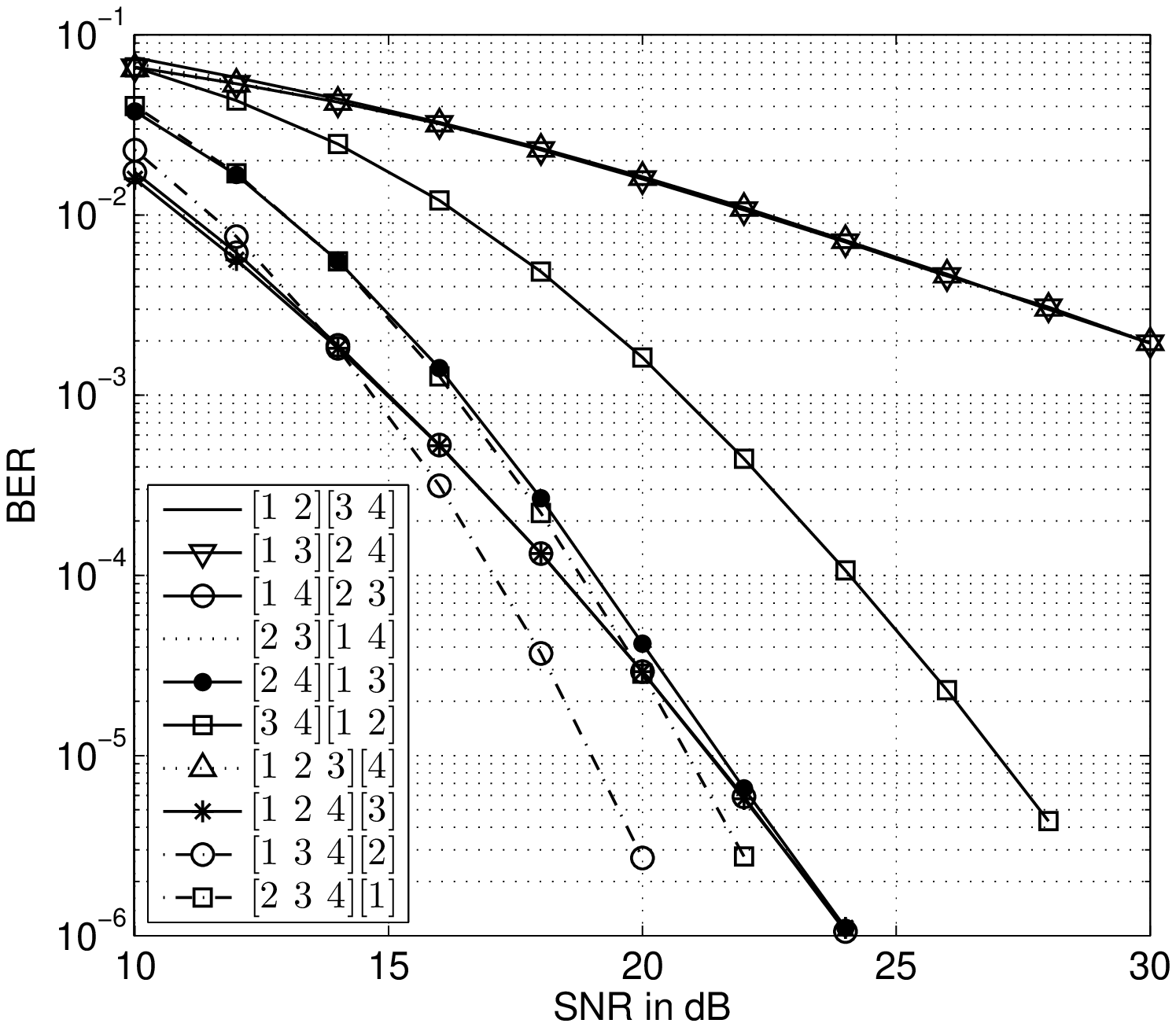} \caption{BER vs. SNR for $4
\times 4$ $S = 4$, $4$-QAM PPMB.} \label{fig:partial_precoding}
\end{figure}
\else
\begin{figure}[!t]
\centering \includegraphics[width =
0.95\linewidth]{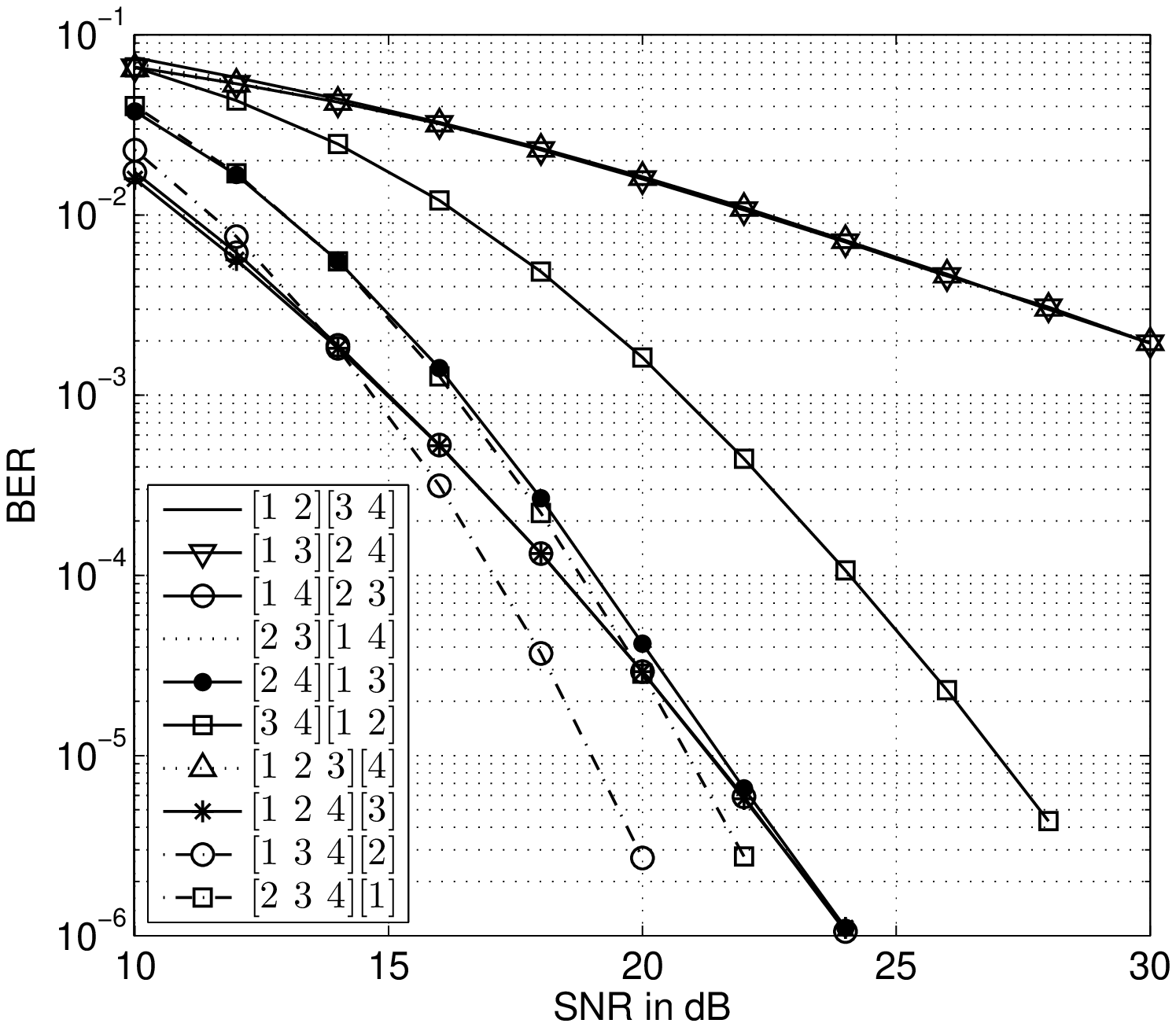} \caption{BER vs. SNR for $4
\times 4$ $S = 4$, $4$-QAM PPMB.} \label{fig:partial_precoding}
\end{figure}
\fi

%% file: conclusion.tex
\section{Conclusion} \label{sec:conclusion}

In this paper, we introduced a precoded beamforming system that
achieves full diversity order. For the analysis, we calculated an
upper bound to the pairwise error probability, assuming that the
receiver decodes the transmitted symbols based on maximum likelihood
decoding. We established several criteria to design the precoder. We
also provided simulation results that support the diversity analysis
of precoded beamforming. We showed in particular fully precoded
multiple beamforming with a proper precoding matrix outperforms
single beamforming at the same system data rate while achieving full
diversity order.